\documentclass[preprint]{elsarticle}

\usepackage{graphics}
\usepackage{graphicx}
\usepackage{natbib}
\usepackage[colorlinks=true]{hyperref}
\usepackage{url}
\usepackage{color}
\usepackage{amsmath,amssymb}
\usepackage[utf8]{inputenc}

\newcommand{\ee}{\end{equation}}
\newcommand{\be}{\begin{equation}}

\newcommand{\eea}{\end{eqnarray}}
\newcommand{\bea}{\begin{eqnarray}}

\begin{document}
\setlength{\parskip}{0.3\baselineskip}

\title{Correlated percolation models of structured habitat in ecology}



\author[l2c,amap]{G\'{e}raldine Huth}
\author[lptmc]{Annick Lesne} 
\author[amap]{Fran\c{c}ois Munoz} 
\author[l2c]{Estelle Pitard}

\address[l2c]{Laboratoire Charles Coulomb, Universit\'{e} Montpellier II and CNRS, 34095 Montpellier, France}
\address[amap]{UM2, UMR AMAP, Bd de la Lironde, TA A-51 / PS2, 34398 Montpellier cedex 5}
\address[lptmc]{Laboratoire de Physique Th\'eorique de la Mati\`ere Condens\'ee UMR CNRS 7600,\\ LPTMC, Tour 12-13/13-23, Bo\^ite 121,
4, Place Jussieu,
75252 Paris Cedex 05, France}
  
  \date{\today}

\vskip 7mm

\begin{abstract}
 Percolation offers acknowledged models of random media when the relevant medium characteristics can be described as a binary feature. However, when considering habitat modeling in ecology,  a  natural constraint comes from nearest-neighbor correlations between the  suitable/unsuitable states of the spatial units forming the habitat.  Such constraints are also relevant in the physics of aggregation where underlying processes may lead to a form of correlated percolation. However, in ecology, the processes leading to habitat correlations are in general not known or very complex. As proposed
 by Hiebeler  [Ecology {\bf 81}, 1629 (2000)], these correlations can be captured in a lattice model by an observable aggregation parameter $q$, supplementing the density $p$  of suitable sites. We investigate this model 
 as an instance of correlated percolation. We analyze the phase diagram of the percolation transition and compute  the cluster size distribution, 
 the pair-connectedness function $C(r)$ and the correlation function $g(r)$. 
  We find that while $g(r)$ displays a power-law decrease associated with long-range correlations in a wide domain of parameter values,  critical properties
  are compatible with the universality
 class of uncorrelated percolation. We contrast the correlation structures obtained respectively for the correlated percolation model and for the Ising model, and show that
 the diversity of habitat configurations generated by the Hiebeler model is richer than the archetypal Ising model. We also find that  emergent structural properties are peculiar to the implemented  algorithm, leading to questioning the notion of a well-defined model of aggregated habitat.
We conclude that the choice of  model and algorithm  have strong consequences on what insights ecological studies can get using such models of species habitat.
\end{abstract}


\maketitle


\section{Introduction}

Much research in ecology is devoted to investigating the dynamics of organisms in space and time, according to their survival, reproduction and dispersal abilities. A challenging issue is to characterize these dynamics under the constraint of a spatially heterogeneous environment, such that some sites allow successful establishment or survival while others do not. The habitat can then be seen as the ensemble of suitable sites, located on a discretized lattice; the other sites being defined as unsuitable, and a priori forbidden to the organism. Ecologists aim at describing the main spatial properties of this habitat based on a minimal number of parameters. They have mainly considered characterizing habitat structure based on its spatial autocorrelation. This focus is related to the fact that: (i) the habitat is generated by other organisms that exhibit diffusive dynamics and aggregated patterns (e.g., a forested vegetation pattern), (ii) an organism that live and move in this habitat has itself limited ability to disperse far away from the sources (e.g. a bird living in the forested vegetation), and thereby also exhibit aggregated spatial structures. In this regard, Hiebeler [12] proposed a model of habitat structure based on a parameter of density, $p$, the proportion of suitable sites in the lattice, and a parameter of aggregation, $q$, the probability to find a suitable site nearby another suitable one. A crucial ingredient of the model is the algorithm that constructs habitat maps with target parameters $(p,q)$.   An overlooked issue in ecology is whether such a convenient and reduced set of parameters allows characterizing any habitat spatial structure in an unambiguous way. We will show that  the answer is clearly no, and the habitat characteristics depend dramatically on the algorithm specifications.The $(p,q)$ parameterization, as well as other versions using longer-range parameters of correlation, are in fact based on an implicit process of habitat generation resulting in the aggregated pattern. But algorithms used to simulate habitats for these parameters perform ad hoc switches of suitable site labels in order to converge to the expected spatial structure. Here we question whether such algorithms based on expected patterns rather than on underlying generating mechanisms allow getting a well-defined typology of habitat, or alternatively generate different results according to the generating algorithm. Whether or not the habitat is defined in an unambiguous way is indeed critical to the analysis of organism dynamics within this habitat.

Percolation theory is one of the most important landmarks in statistical physics, in relation with the theory of critical phenomena. The basic percolation model, namely uncorrelated site percolation, only prescribes habitat density $p$, such as each site is suitable independently from the other sites. The model displays a critical transition at a value $p_c$, above which the lattice is spanned by an infinite cluster (connected set) of suitable sites. The threshold value $p_c$ has been determined for a variety of infinite lattices, with $p_c \simeq 0.592$ in a square lattice \cite{bookStauffer}. Models of correlated percolation conversely consider that the probability of a site being suitable depends on the status of its neighbors \cite{ConiglioFierro2008}. 
Another archetypal model of statistical physics, the Ising model, is based on local particle interactions resulting in patterns of spatially correlated structures \cite{Ising,Baxter}. The spatial properties of this model are well-known and can thereby serve as a reference to understand the structure of correlated habitat. However the magnetic field $h$
 and the interaction parameter $J$ of the Ising model are very difficult to calibrate in the context of an ecological habitat depending on many underlying dynamical processes. That is why ecologists have still resorted to describe habitats based on resulting habitat structure rather than based on generating mechanisms. We therefore wish to assess the relationship between models of correlated percolation of physics with the models of correlated habitat of ecologists. We focus on correlated percolation models prescribed by local rules, namely correlations between nearest neighbors, to comply with the minimal parameterization $(p,q)$ of Hiebeler.
 
 The Hiebeler model is hence a new model of correlated percolation that has never been investigated in the past in the light of statistical physics. Such a model could also be of interest in materials physics for the design of porous media for instance \cite{Bejan}.
We will show that the choice of the $(p,q)$ parameterization, though appealing for ecologists, is not unequivocal and that the consequences are critical for the understanding of organism spatial dynamics in structured habitats. Our demonstration is based on a comparison of the properties of the Hiebeler model with the Ising model, and on an assessment of the sensitivity of these properties to algorithmic choices. In Section II, we formulate in physical terms the nearest-neighbor model of aggregation introduced on ecological grounds by Hiebeler [12]. Section III is devoted to the analysis of the Hiebeler  model in the light of statistical physics and percolation theory. We investigate its phase diagram and compute its  spatial properties. In Section IV we show that a mapping is possible between the Hiebeler model and Ising model, though being incomplete and sometimes equivocal. This allows highlighting differences between the two models, the Hiebeler model showing more diversity in configuration space that its physical counterpart. In  Section V, we discuss the relevance of the Hiebeler model for modeling ecological habitats. In particular we discuss how dynamical models with local dispersion rules can be most conveniently studied on a Hiebeler type of habitat, with the aggregation parameter $q$ being the fundamental quantity necessary to assess the survival of organisms on such substrates.

\section{The nearest-neighbor aggregation model of Hiebeler}


\subsection{Pairwise conditional probabilities}

The first prescription of the model is an  homogeneous probability of site suitability: \be {\rm Prob} (X_{\vec{r}}=1)=p\ee where $\vec{r}$ labels the  sites of a square lattice
and $X_{\vec{r}}$ is the Boolean random variable describing the status of the site $\vec{r}$:  $X_{\vec{r}}=1$ if the site is suitable, otherwise $X_{\vec{r}}=0$.
In contrast  to  uncorrelated percolation, the suitability of sites is now correlated according to the prescription of conditional probabilities between nearest neighbors:
\be  \label{eq:q} {\rm Prob} (X_{\vec{r}+\vec{u} } =1 \;|\; X_{\vec{r}}=1)=q\ee
for any $\vec{u}\in {\cal U}=\{ (0,1), (0, -1), (1, 0), (-1, 0)\}$.
 $q$ is termed the {\it aggregation parameter}.
It is straightforward to derive the other nearest-neighbor conditional probabilities, in particular
\be  {\rm Prob} (X_{\vec{r}+\vec{u}}=0 \;|\; X_{\vec{r}}=0)={1-2p+pq\over 1-p}.\ee
Requiring the non-negativity of this probability yields  the following condition
\be q\geq 2-1/p.\ee
Therefore, when considering the behavior of the model at varying ($p$,$q$) values, the region defined by $q < 2-1/p$ will be
an excluded region \cite{Hiebeler2000}.

Based  only on these constraints, the model  does not prescribe all the moments of  the state variable $X_{\vec{r}}$, nor even the second moment for any pair of sites. Both  spatial and statistical closures are required to get a full knowledge of the global distribution.

In the uncorrelated percolation model,
 ${\rm Prob} (X_{\vec{r}}=1 \;|\; X_{\vec{r}+\vec{u}})=p$ 
  independently of the suitability status of $X_{\vec{r}+\vec{u}}$.
Uncorrelated percolation is  here recovered when $q=p$, with a  site-percolation threshold $p_c^*\approx 0.59$ on a square lattice  \cite{bookStauffer, LL}. 



\subsection{Hiebeler's algorithm and some alternatives}

Given the parameters $p$ and $q$, the algorithm  proposed by Hiebeler  \cite{Hiebeler2000}  for generating binary configurations  of suitable/unsuitable habitat on a square lattice proceeds in two steps. 
In the first step, each site of the lattice is labelled suitable with a probability $p$ independently of the other sites. In the second step,  a
site is chosen at random and its state is switched 
if this change induces a reduction in the difference between the desired and the observed ($obs$) joint probabilities,
measured via the following cost function $D$: 
\be D = \sum_{\langle s, s'\rangle, i,j} \,| \, p_{ij}^{obs}(s,s') - p_{ij} (s,s')\,| \, \ee 
where the sum runs over the pairs $\langle s, s'\rangle$ of nearest neighbors, and over  states $i$ and $j$ taking the values $0$ or $1$.
This step is applied repeatedly, until a given maximum number of iterations is achieved, or when the difference between the desired and observed probabilities reaches a
low-enough  predetermined threshold (we chose the default threshold equal to 0.001 as used in  \cite{Hiebeler2000}). This procedure implies that the final values of both $p$ and $q$ can be slightly different 
from their target values.
As mentioned in  \cite{Hiebeler2000}, the convergence is hard to achieve
 for some pairs $(p,q)$, typically in the situation of low suitable habitat and large clustering
 ($p$  small  and $q$  large). In this case, the difference between the initial and the desired values of the parameters  is typically larger than in other regions of parameter space.


An important question raised by this algorithm is to understand to what extent the local prescriptions  $p$ and $q$  control the overall spatial structure of the simulated habitat.
To investigate this question, we compared Hiebeler's algorithm with two other  methods in order to determine whether some implicit constraint in the algorithm affects the global properties of the system.
For these two alternative algorithms, the first step is still the same  and consists in a random independent filling of the lattice sites with the probability $p$. 
In a first variant,  one picks randomly a pair of sites (instead of randomly picking one site in the lattice)  and tries to exchange their suitability values. If this exchange reduces the value of the cost function $D$ the new configuration is accepted, otherwise one picks another pair of sites, until $D$ reaches the threshold value.
In this procedure, the density of suitable sites remains constant.
This algorithm was already proposed  in  \cite{ThomsonEllner2003}.
The second  variant   consists in picking only pairs of neighboring sites, instead of any pair of sites on the lattice.
In both variants the convergence  is speeded up.
Another variant was proposed by Hiebeler in  \cite{Hiebeler2007}, but the convergence of the algorithm  remains slow for some range of parameters. The following Section~III presents how  the global properties of the model are sensitive to such changes in the algorithm.

%
%

\section{Percolation transition and correlation properties of the Hiebeler model}

We used Hiebeler's algorithm and  the alternative algorithms presented above to simulate
matrices of binary habitats of size 1000$\times$1000.
This choice  of  $L=1000$ corresponds to a compromise between long computation times for large lattices
 and a more important variability between runs for smaller lattice sizes. The results were  checked and confirmed by implementing  simulation
  with a  larger lattice   of size
 2500$\times$2500 when it was tractable.  
In several cases where finite-size effects were too large to give reliable results, especially for estimating critical exponents, we also performed simulation
 on 10000$\times$10000 lattices. Unless explicitly mentioned, results presented below are  based on the original Hiebeler's algorithm with $L=1000$.

\subsection{Percolation transition and phase diagram}

\subsubsection{Zero-one percolation transition}

\begin{figure}[h]
\centering
\includegraphics[scale=0.5]{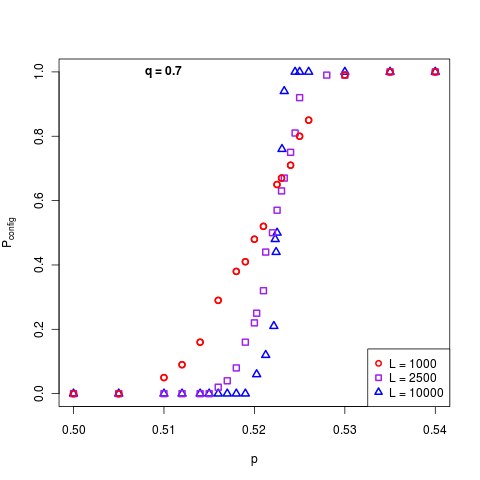}
\caption{(Color online) Numerical support of the zero-one law for correlated percolation. The variation of the probability $P_{config}(L, p, q)$ that a configuration
contains a spanning cluster as a function of $p$ is shown for different values of  
the lattice linear size $L$ and a fixed value $q=0.7$. The curves cross approximately at $p_c(q=0.7)$.
They go steeper as $L$ increases, which supports the convergence to a step-like behavior in $p_c$ in an infinite lattice $L=\infty$, 
in agreement with a zero-one law. This probability $P_{config}(L, p, q)$ 
is estimated by counting the number of percolating configurations among 100 realizations of the lattice, by using the Hoshen-Kopelman algorithm  \cite{HoshenKopelman}.
}
\label{default1}
\end{figure}

A zero-one behavior has been demonstrated in the case of uncorrelated percolation, stating that either almost all or almost none of the configurations of the square lattice percolate,
according to the value of $p$  \cite{zero-one-percol}.  The mathematical proof is based on the fact that the existence of an infinite cluster is a tail event, that is, it is not affected by a change in the state of a finite number of cells. Hence, the standard zero-one law for uncorrelated sequences of random variables (here, the binary variables describing cells suitability) ensures that the probability of this event is either 0 or 1  \cite{Bartfai}.
However, the extension of the zero-one law to correlated sequences of random variables is not straightforward  \cite{sucheston}.  Computing the probability
 $P_{config}(L, p, q)$  that a configuration
contains a spanning cluster as a function of $p$  at fixed $q$  and increasing  $L$ numerically supports   an all-or-none behavior
  in  infinite lattices (Fig.~\ref{default1}).
Although this result has not the status of a rigorous mathematical proof, it gives a strong clue that a zero-one law holds in the case of correlated percolation.
We underline here that $P_{config}$ is a probability in the space of lattice configurations. 
Accordingly, a zero-one behavior
 would ensure  that  almost all configurations (for $p>p_c$) or almost none (for $p<p_c$) percolate.  This  typicality of  individual realizations is a pre-requisite to a numerical study, since it makes possible
  to detect a {\it bona fide} transition on a single realization of the system.

%
%

\subsubsection{Percolation probability}

By analogy with uncorrelated percolation, the order parameter of correlated percolation is the percolation probability $P(p,q)$, 
defined as  the fraction of sites belonging to the infinite cluster (equal to 0 when there is no infinite cluster). 
In the case of uncorrelated percolation, percolation probability obeys a scaling law $P(p)\sim (p-p_c)^{\beta}$ above the threshold, and $P(p=1)=1$. 
The associated exponent is universal insofar as it depends only on the dimension, but not on the geometry of the lattice, in contrast to the lattice-dependent threshold value $p_c$. Percolation probability should not be confused with the above-mentioned probability $P_{config}$ that a configuration percolates (i.e. the fraction of percolating configurations).


The transition line $q_c(p)$, separating the percolating phase from the non-percolating phase in the $(p,q)$ plane, can be defined as follows (with the condition $q\geq 2-1/p$, mentioned earlier):
\be 
q_c(p)=\inf\{q, P(p,q)>0\}\ee
or equivalently  as a transition line $p_c(q)$ where
\be p_c(q)=\inf\{p, P(p,q)>0\}.\ee
The uncorrelated percolation threshold $p_c^*$ is recovered for $p=q$.

\begin{figure}[htbp]
\begin{center}
\includegraphics[width=0.7\columnwidth]{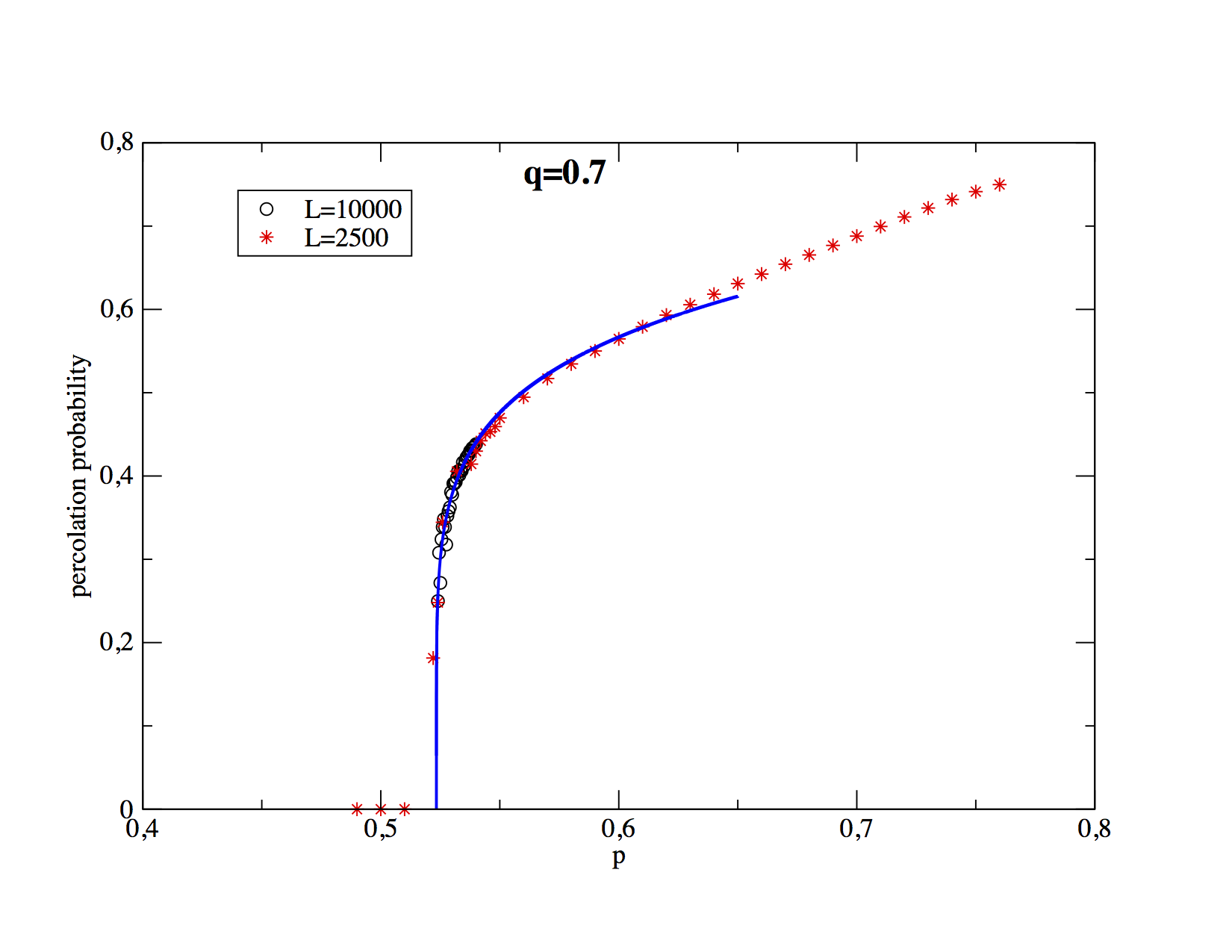}
\caption{(Color online) \ Variation of the percolation probability $P(p,q)$ as a function of $p$, for a fixed value $q=0.7$.
 The curve is in agreement with a convergence to a power-law behavior $P(p,q)\sim (p-p_c(q))^{\beta(q)}$   in infinite lattices.  In the simulation, the lattice linear size is $L = 2500$, and the points near the percolation threshold were
 obtained from a simulation on a larger lattice with $L=10000$.  The best fit gives $\beta \approx 0.16$, while 
   $\beta=5/36 \approx 0.139$  for uncorrelated percolation. The value of $p_c(q=0.7) \simeq 0.523$ can be read from the horizontal axis. 
   }
\label{default2}
\end{center}
\end{figure}

We show on Fig.~\ref{default2} the numerical estimation of the percolation probability
for $q=0.7$, but we have also performed simulations for a wide range of $q$ values for which the percolation transition occurs. One finds near the threshold $p_c(q)$ a scaling law
$P(p,q)\sim [p-p_c(q=0.7)]^{\beta(q=0.7)}$. The value of the exponent $\beta$ 
is obtained by a power law fit. This requires 
 a large lattice $L=10000$ to avoid too large finite-size effects.
For $q=0.7$, we find $\beta \approx 0.16$, and for different values of $q$, the results vary from
0.12 to 0.16. This is compatible, within error bars, with the exponent given by uncorrelated percolation,  $\beta=5/36 \approx 0.139$.
We checked that a similar  error bar range  (containing the exact value) is obtained  with an uncorrelated percolation  model on  a lattice of the same size. It is indeed  well known 
 \cite{bookStauffer} that very large lattice sizes are needed in order to 
 accurately determine the critical exponents. Therefore,
our results are consistent with a  universal exponent  $\beta=5/36$ for all values of $q$. 

\subsubsection{Phase diagram}

\begin{figure}[htbp]
\begin{center}
\includegraphics[width=0.5\columnwidth]{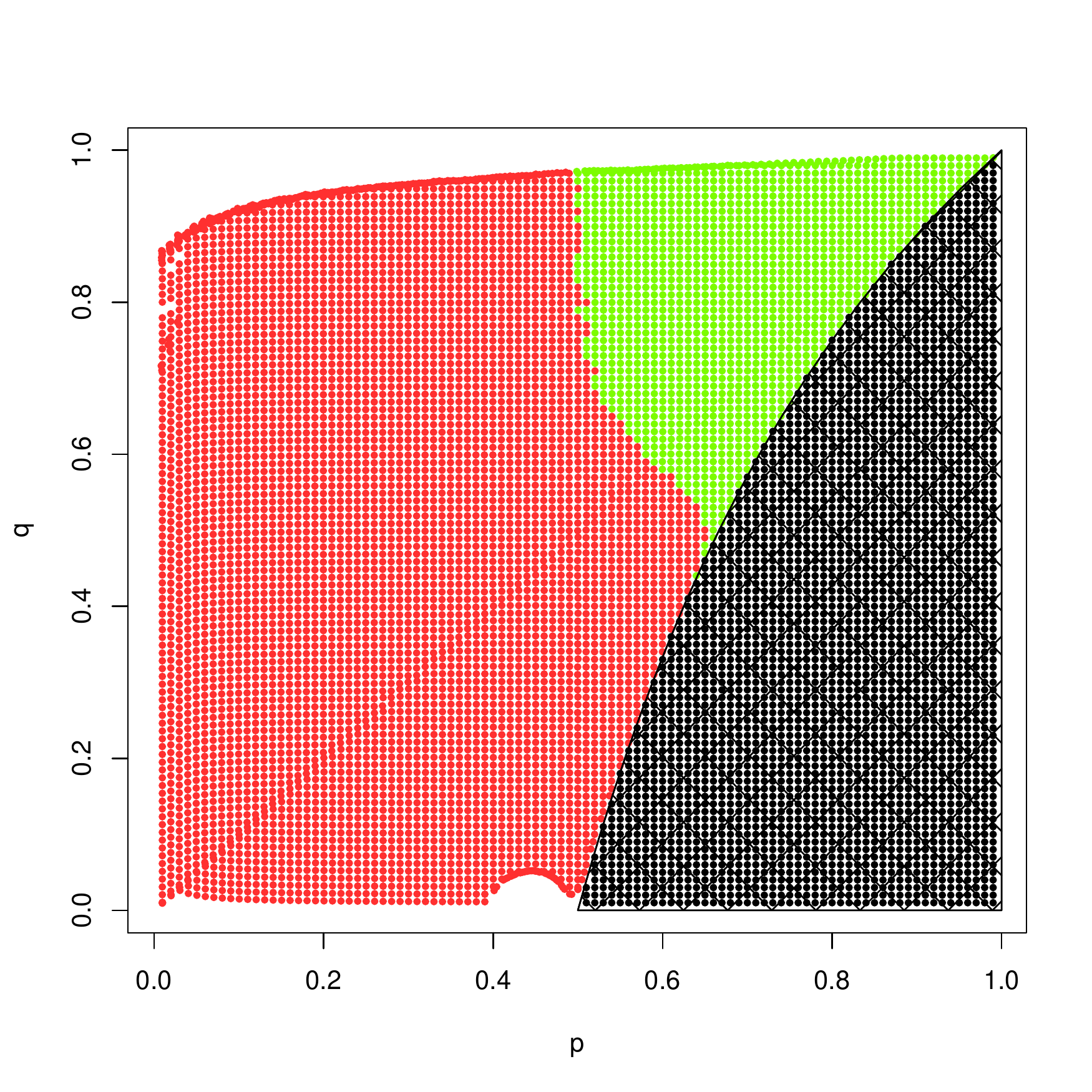}
\caption{(Color online) \ Phase diagram for the original Hiebeler's algorithm in the $(p,q)$ plane. Each point corresponds 
to a different simulation. In green/light grey are represented percolating configurations 
and in red/dark grey non percolating configurations. 
The black zone corresponds to excluded parameter pairs (when ${q\leq 2-1/p}$).
 The network linear size is $L=1000$ which corresponds to stable results  (independent of $L$) . 
 }
\label{default3}
\end{center}
\end{figure}

Fig.~\ref{default3} represents the phase diagram of Hiebeler's simulated habitats in the $(p,q)$ plane
in terms of percolating or non-percolating phases. Because of the finite size of the configurations, the percolation threshold
depends on the size $L$ of the lattice.
It also  varies from one configuration to another and it can be  difficult to extrapolate with precision its (deterministic) value $p_c^*(q)$ in an infinite lattice. However, 
we checked (see the colored/filled regions) that one can obtain stable results for the sizes studied, except  in some  regions of parameters (white/empty regions) 
where the Hiebeler algorithm is too slow to converge to the desired $(p,q)$ values.

\begin{figure}[htbp]
\begin{center}
\includegraphics[width=0.5\columnwidth]{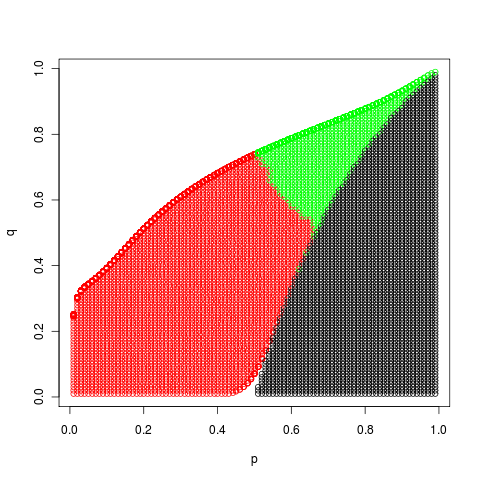}
\caption{(Color online) \ Phase diagram in the $(p,q)$ plane for the modified algorithm involving only permutations
of neighboring sites: the empty zone corresponds to situations where the convergence of the algorithm
is very slow. In other regions the phase diagram is similar to the original Hiebeler's algorithm. 
 The network linear size is $L=1000$. }
\label{default4}
\end{center}
\end{figure}
\begin{figure}[htbp]
\begin{center}
\includegraphics[width=0.5\columnwidth]{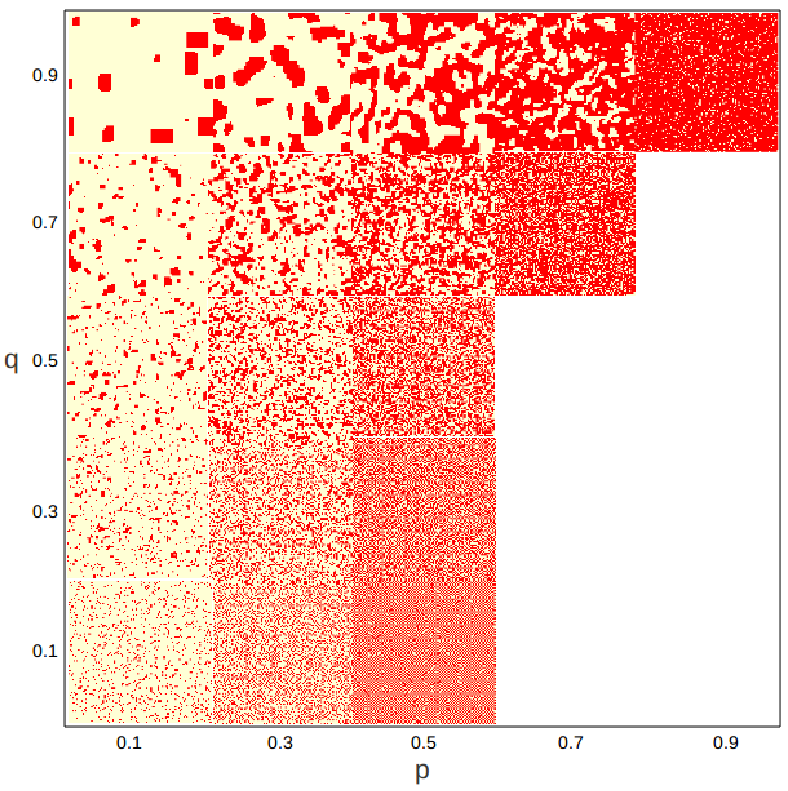}
\caption{(Color online) \ Phase diagram in the $(p,q)$ plane using Hiebeler's algorithm.
 The blocks display the  typical configurations observed for 18 admissible pairs of values $(p,q)$. In red/dark grey  are the suitable sites and
 in yellow/light grey the unsuitable sites: we will keep the same color code in all the figures. The empty zone corresponds to the inaccessible range of parameters $q < 1 -1/p$.}
\label{default5}
\end{center}
\end{figure}

The transition line in the $(p,q)$ plane  between the percolating phase and the  non-percolating phase  remains the same ---up to convergence limitations--- when
one changes the algorithm generating the configurations. This is illustrated in Fig.~\ref{default4} which corresponds to the modified algorithm where one proceeds by state permutations of neighboring sites.
In this latter case, the convergence is slower than for Hiebeler's algorithm, which is rather intuitive, considering that this other algorithm introduces space correlations (because of the choice of pairs of neighbors) compared to a single random choice of site. 
Typical configurations illustrating this percolating transition are presented in Fig.~\ref{default5}.


\subsection{Cluster statistics and correlation functions}

As in standard  percolation theory, configurations of the above correlated percolation model -using Hiebeler's algorithm- can be spatially characterized  by  cluster statistics and spatial correlation functions.


\subsubsection{Size distribution of finite clusters}

The size distribution of finite clusters (connected sets of suitable sites) is of great ecological relevance to characterize the patterns of insularity and fragmentation that constrain biodiversity dynamics. It recovers the general belief that not only habitat loss, but also habitat fragmentation and connectedness, matters to conservation issues  \cite{Fahrig2003}.


Coniglio  and collaborators  \cite{ConiglioRusso1979, ConiglioFierro2008} presented percolation theory as the generic framework to investigate the distribution of clusters  given the distribution of the constitutive elements (here suitable sites), with the underlying idea that
geometrical clusters  capture the physical properties of the substrate.

 A relevant quantity is    the distribution  $P(s)$  of clusters  of suitable sites not belonging to the infinite cluster and belonging to a cluster of size $s$.
In the case of uncorrelated percolation, one has
$P(s) \sim s^{-\tau}e^{-s/\zeta(p)}$. A power-law behavior is observed only at the percolation transition, where the coherence length $\zeta(p)$ diverges according to the scaling law $\zeta(p)\sim (p-p_c)^{-\nu}$ with $\nu=4/3$, and $\tau=187/91 \approx 2.055$ \cite{bookStauffer}.

\begin{figure}[htbp]
\begin{center}
\includegraphics[width=0.6\columnwidth]{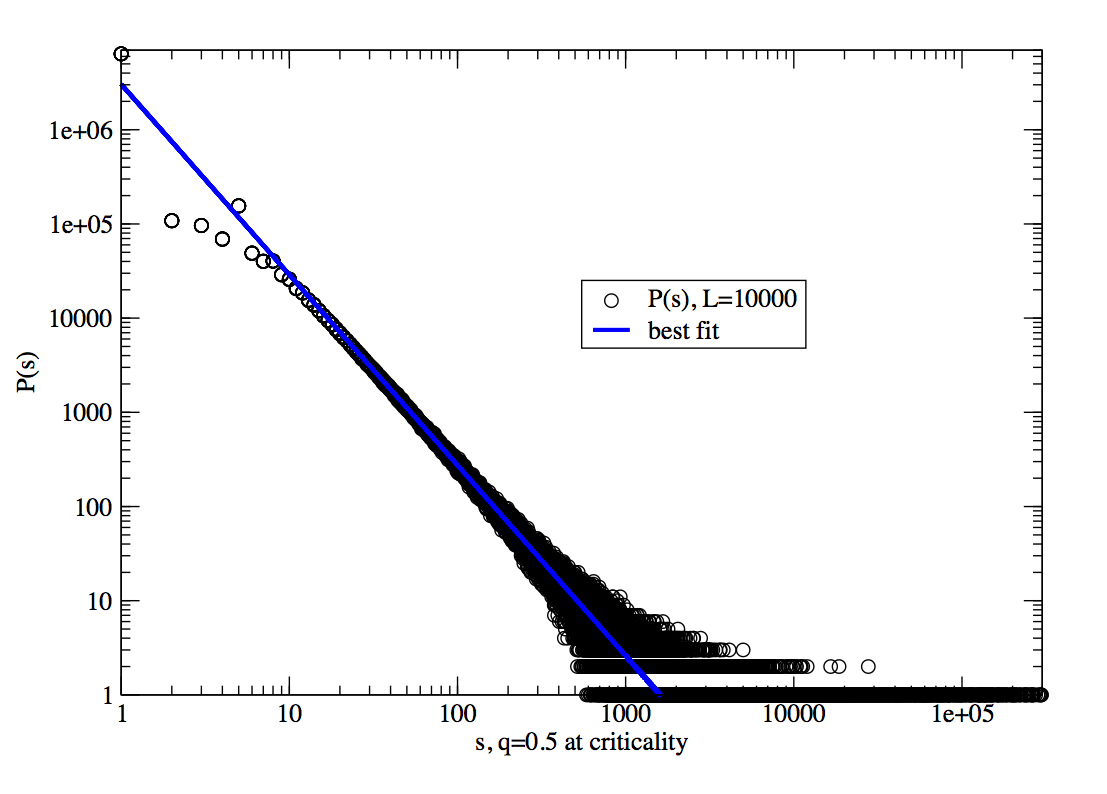}
\includegraphics[width=0.6\columnwidth]{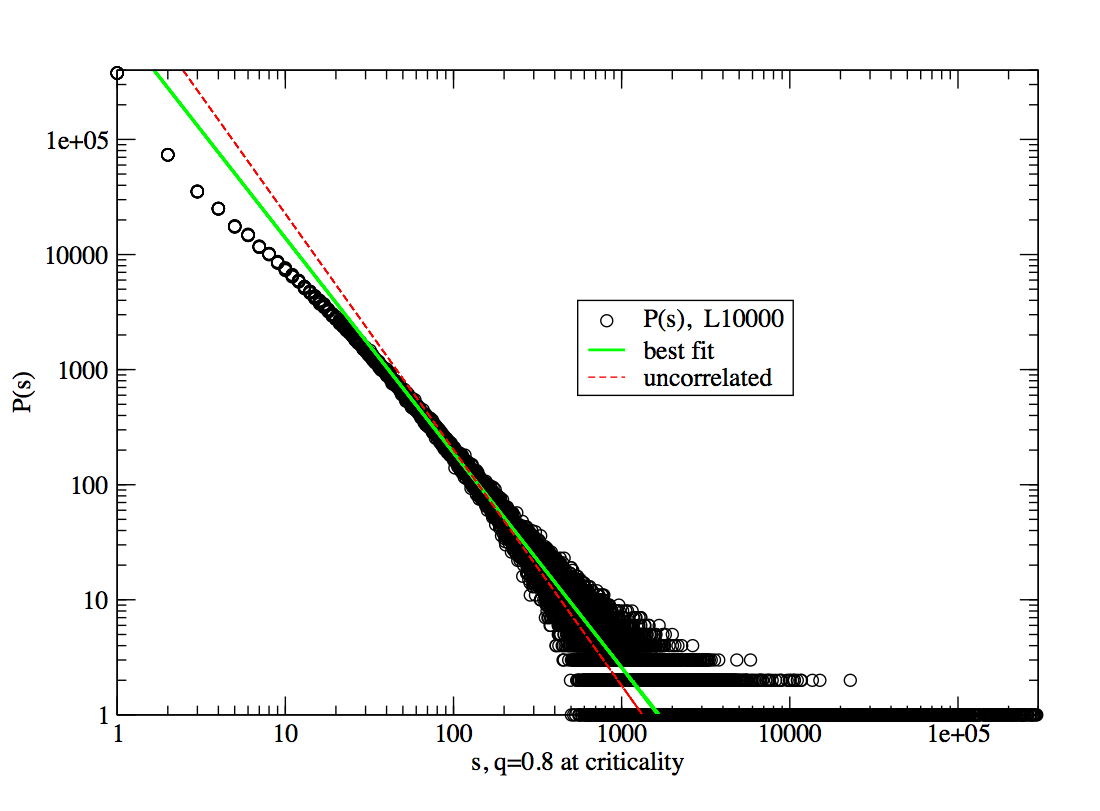}
\caption{(Color online) \ Log-log plot of the cluster size distribution $P(s)$ ---up to a normalization factor--- in the Hiebeler's model for (top) $q=0.5$ and (bottom) $q=0.8$ at criticality.
 The lattice size is $L=10000$ and the critical point was accurately determined for the same lattice size.
At large $s$ a power law $P(s)\sim \tau^{-s}$ is observed, with an exponent $\tau=2.02$  for $q=0.5$  and 
$\tau=1.86$ for $q=0.8$  very close to the uncorrelated percolation exponent  value $\tau=187/91$.
 }
\label{default6}
\end{center}
\end{figure}

\begin{figure}[htbp]
\begin{center}
\includegraphics[width=0.6\columnwidth]{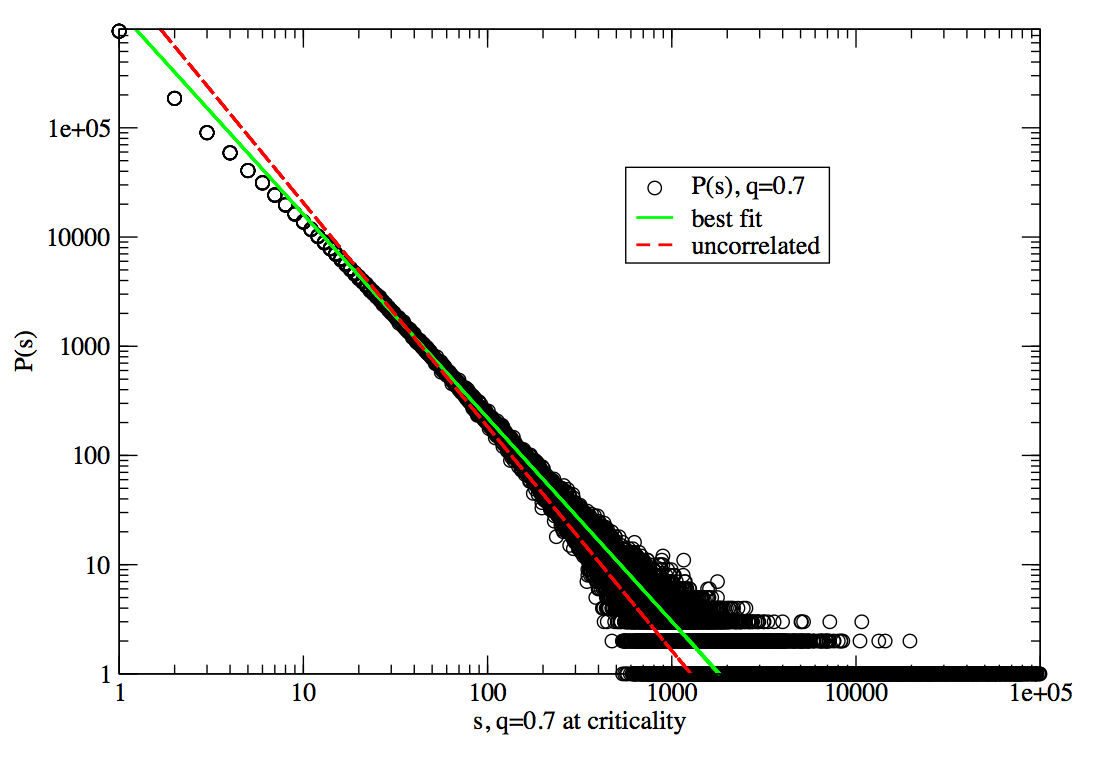}
\includegraphics[width=0.6\columnwidth]{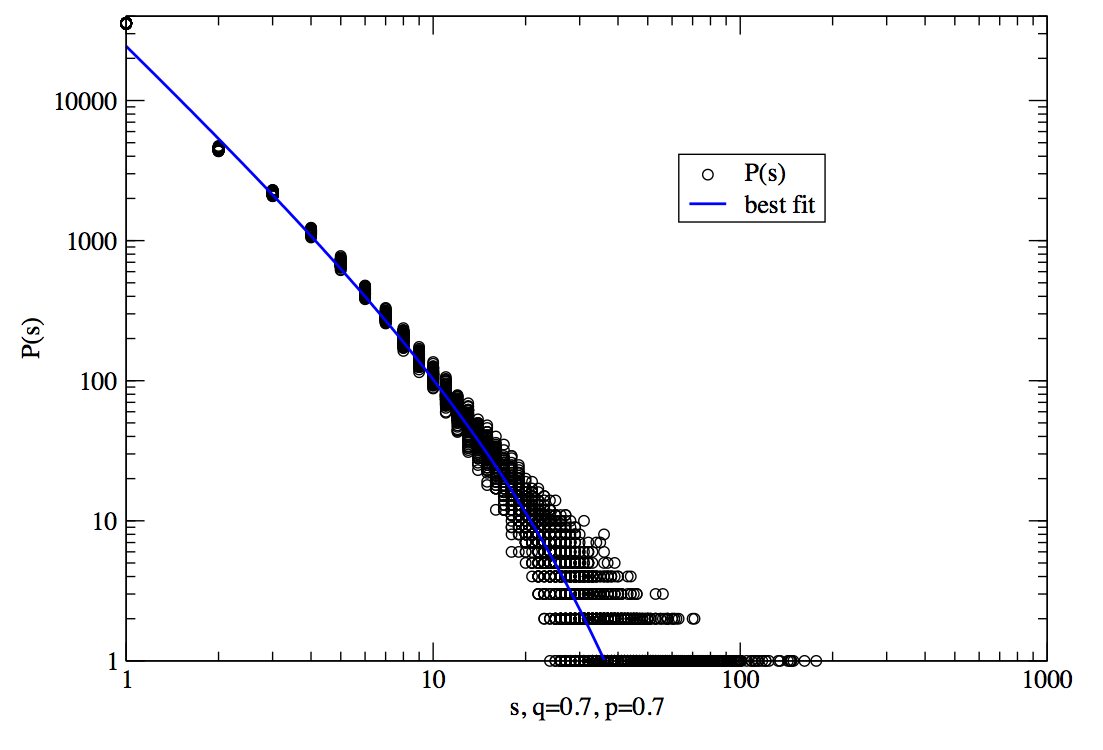}
\caption{(Color online) \ Log-log plot of the cluster size distribution $P(s)$ ---up to a normalization factor---  in the Hiebeler's model for $q=0.7$  at (top) criticality $p=p_c(q=0.7)$ and  (bottom) for $p=0.7$.
 At large $s$ a power law $P(s)\sim \tau^{-s}$  is observed at criticality, again with an exponent $\tau$ compatible within
the error bars with the uncorrelated percolation exponent  (numerical value $\tau\approx 1.92$ for  $q=0.7$). Away from criticality, an exponential part
superimposes on the power-law: for $p=0.7$, the best fit gives $P(s) \sim e^{-s/13.5} s^{-2.09}$. }
\label{default7}
\end{center}
\end{figure}

We  calculated the  cluster size distribution $P(s)$ for the Hiebeler's model at the critical point $p=p_c(q)$ for various values of $q$.
At the percolation threshold, the distribution is a power-law $P(s) \sim s^{-\tau(q)}$, with an exponent $\tau(q)$ which varies between 1.86 and 2.02 when $q$ varies. In all cases the  value $\tau=187/91$ for uncorrelated percolation lies within  the error bars obtained through the fitting procedure
 (Figs.~\ref{default6} and \ref{default7}). 
 We checked that a similar  error bar range  (containing the exact value) is obtained  with the  uncorrelated percolation  model at criticality  ($p_c=q_c=0.592..$)
 on  a lattice of the same size.

Outside the threshold, $P(s)$ is  not a power law. It can be fitted by $s^{-\tau(p,q)}e^{-s/\zeta(p,q)}$ where  the value of $\tau(p,q)$  is again consistent with the uncorrelated  percolation value. It is not our objective here to characterize precisely the critical behavior of $\zeta(p,q)$, as it would require extensive simulation as a numerical project in itself.
As far as the critical exponent $\tau$ is concerned, we can say that its value is compatible with the value of uncorrelated correlation.

  In the case of correlated percolation, results available in the literature are scarce.
   A power-law decay is claimed to be found only at the transition  \cite{VandenBerg2011}. Our results tend to confirm this property, together with the fact that critical exponents do not seem to change drastically between uncorrelated and correlated percolation.


\subsubsection{Pair connectedness function $C(r)$}\label{subsec:corr-fn}

The  pair connectedness function $C(r)$  is defined as the probability that a site at distance $r$ from a suitable origin is suitable and belongs to the same {\it finite} cluster (the infinite cluster has to be  removed from the computation)  \cite{ConiglioFierro2008, ConiglioRusso1979}. Statistical isotropy of the model ensures that $C(r)$ depends only on the modulus $r$.

The typical behavior of this correlation function is
 \be
 C(r)\sim {e^{-r/\zeta_{corr}} \over r^{d-2+\eta}}\ee
 where $\zeta_{corr}$ defines another coherence length and characteristic cluster linear size  (a priori different from $\zeta(p,q)$). 
 For uncorrelated percolation, its has been shown \cite{bookStauffer}
 that $\eta=2\beta/\nu$ which yields $\eta=5/24\approx 0.208333$ in $d=2$ (exponent values are rational in $d=2$, according to conformal invariance arguments). 

\begin{figure}[htbp]
\begin{center}
\includegraphics[width=0.5\columnwidth]{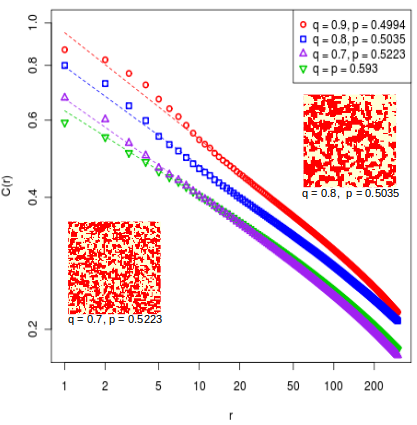}
\caption{(Color online) \ Log-log plot of $C(r)$ for various values of  $q$ at the percolation threshold consistent with a power-law decrease $C(r)\sim r^{-\eta}$. The dashed lines are the best fits, giving  the value of $\eta$; corresponding maps are shown, for $L=2500$.}
\label{default8}
\end{center}
\end{figure}

\vskip 3mm
{\it (i) At the percolation threshold}
\vskip 3mm
For $p=q$ (uncorrelated percolation) we found $\eta$ = 0.1941,  where the small discrepancy with the exact  value can be attributed to finite-size effects.
 The values we found for correlated percolation
  deviate from the uncorrelated percolation value when $q$ increases: for $q=0.593, 0.7, 0.8, 0.9$ the values of $\eta$ were  respectively
  $0.194, 0.216, 0.224, 0.2417$ (Fig.~\ref{default8}).  
  This trend would mean that  $C(r)$ seem to decrease more quickly as a function of the distance when the aggregation parameter $q$ is larger.  
  The statistical analysis has been made using  between 30 and 100 different configurations of a lattice of size
   $L=2500$. Even with such statistics, the results may not be completely converged.
   However, the discrepancy between the observed exponents and those of uncorrelated percolation still lies within the error bars and the range of finite-size fluctuations. It thus  seems reasonable to state that Hiebeler's model lies in the same universality class as uncorrelated percolation.

\begin{figure}[htbp]
\begin{center}
\includegraphics[width=0.5\columnwidth]{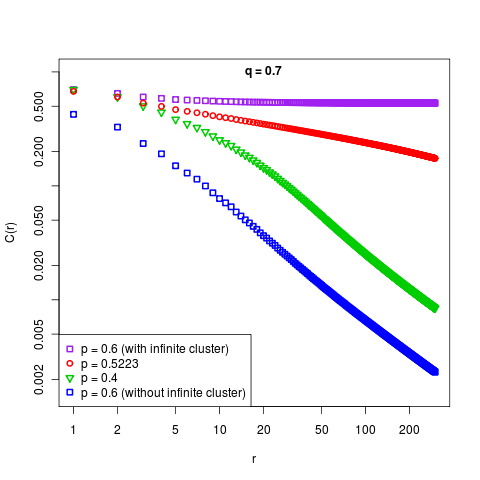}
\caption{(Color online) \ Log-log plot of $C(r)$ for a fixed value of $q=0.7$, at the threshold ($p=p_c(q=0.7) = 0.5223$) 
and outside the threshold (for $p$ = 0.6 and $p$ = 0.4). The upper curve corresponds to the computation of $C(r)$ for $p=0.6$  taking into account the infinite cluster
as well as the other clusters: in this case there is a saturation at large $r$, which is not observed when one removes the dominant infinite cluster from the analysis (bottom curve).}
\label{default9}
\end{center}
\end{figure}

\vskip 3mm
{\it  (ii) Outside the percolation threshold}
\vskip 3mm
Considering only finite clusters, we studied the correlation function outside the percolated threshold and found a steeper decay of $C(r)$ than at the threshold (Fig.~\ref{default9}), in agreement with equation (8), though it is not our purpose here to study systematically $\zeta_{corr}$.


For non-percolating situations ($p<p_c(q)$) this result is expected because the clusters are smaller than at the threshold. Above the threshold ($p>p_c(q)$), 
the decorrelation is steeper because only a few small isolated cluster remain apart from  the dominant cluster (recall that $C(r)$ is computed without taking into account the infinite cluster).
Above the percolation threshold, when the dominant infinite cluster is considered in the calculation, $C(r)$ converges   to a value close to $p$: 
the majority of sites belongs to the infinite cluster (Fig.~\ref{default9}).
We compared the results found using different algorithms for generating the configurations (permutations of pairs of neighboring sites, or permutation of any pairs of sites)
 and found that  the overall shape of
$C(r)$ is not affected.

\vskip 3mm



\subsubsection{Correlation function  $g(r)$}


We have also calculated numerically the correlation function $g(r)$, defined as the probability that a site at a distance $r$ from 
a suitable site is also suitable, independently of the cluster it belongs to.
Contrarily to $C(r)$ which is a more local quantity, $g(r)$ is more suitable to characterize the overall structure of the habitat.
For $r$ = 1, we recover $g(1) = q$, the probability that a neighboring site of an suitable site is suitable. When $r$ increases, we observe a
  decorrelation and $g(r)$ tends to $p$.

\begin{figure}[htbp]
\begin{center}
\includegraphics[width=0.5\columnwidth]{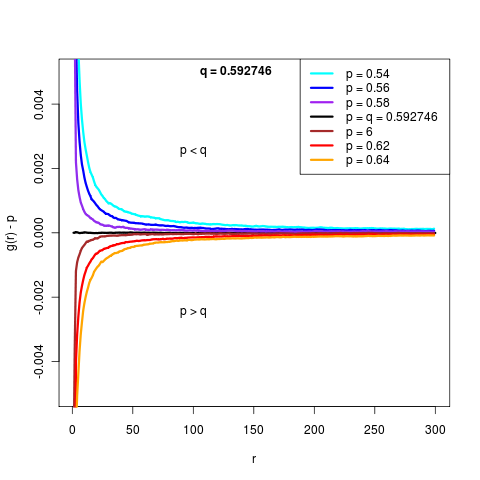}
\caption{(Color online) \ Plot of $g(r)-p$ for the Hiebeler's model for various values of $p$ fixed $q=0,592746$. When $p=q$ one recovers the uncorrelated percolation at criticality, with $g(r)=p$.
}
\label{default11}
\end{center}
\end{figure}

\begin{figure}[htbp]
\begin{center}
\includegraphics[width=0.5\columnwidth]{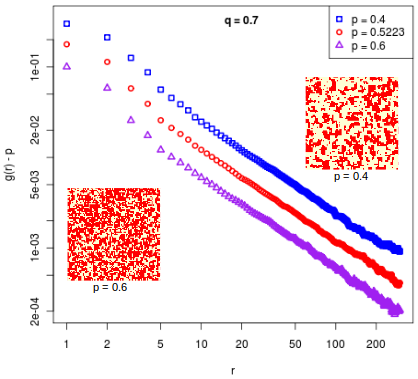}
\includegraphics[width=0.5\columnwidth]{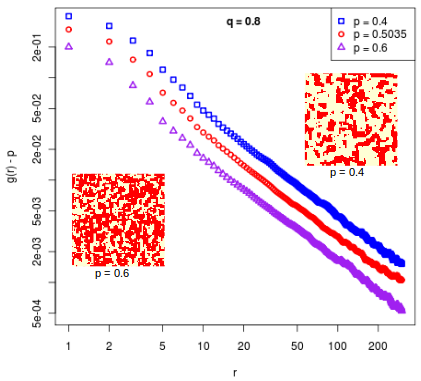}
\caption{(Color online) \ Log-log plot of $g(r)-p$ for the  Hiebeler's model, for (top) $q=0.7$ and (bottom) $q=0.8$. 
When the percolation threshold is crossed by varying the value of
$p$, nothing particular happens in the shape of $g(r)-p$. The maps are shown for $L=2500$.
}
\label{default13}
\end{center}
\end{figure}

Our numerical simulation shows that 
$g(r)-p$ is  always close to a power law (Fig.  \ref{default13}) with an exponent close to 1 at the percolation threshold, i.e. $g(r)-p\sim 1/r$.
Moreover we find no qualitative differences  between percolating and non percolating situations: the exponent remains close to 1
 when varying $p$ at a fixed value of $q$, and even for different values of $q$. 




\begin{figure}[htbp]
\begin{center}
\includegraphics[width=0.5\columnwidth]{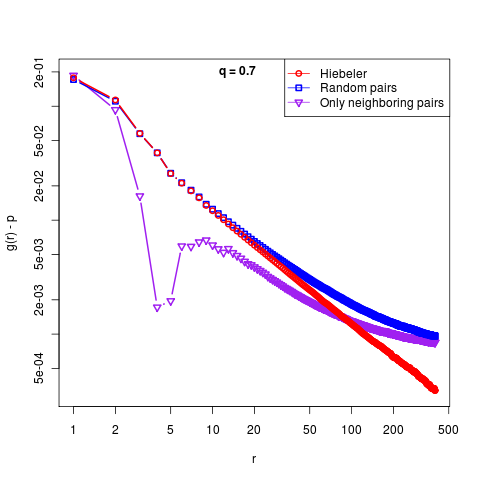}
\caption{(Color online) \ Changes in the  log-log plot of $g(r)-p$ according to the algorithms used to generate the ($p$,$q$)-prescribed habitats, for $q=0.7$ at the percolation threshold. The result with the original Hiebeler's algorithm differs markedly at small and large distances from other choices of algorithms.
}
\label{default14}
\end{center}
\end{figure}

We checked the robustness of this result when one changes the algorithm used for habitat generation (Fig.~\ref{default14}). Strikingly, there are noticeable differences in $g(r)-p$ (we checked, that for the quantity $C(r)$, the changes are minor). When one uses permutations of pairs of neighboring sites, $g(r)-p$  decreases more slowly than a power law at large $r$. At small distances,
 there exists a noticeable dip in the spatial correlations. Thus, the structure of the habitat is strongly changed at small and large distances.
  Hence Hiebeler's original algorithm has its own special properties and cannot be considered as a universal scheme for correlated habitats.
 

\section{Relation to the Ising model}

The Hiebeler's model  locally prescribes correlation constraints.
It is natural to compare this framework with the Ising's model, which is the simplest model of short-range interactions in magnetism and a reference model in statistical physics \cite{Ising, Baxter}.
While nearest-neighbor pair correlations are prescribed in the Hiebeler's model, they arise from the prescription of nearest-neighbor  interactions (interaction parameter $J$) and the external magnetic field $h$  in the Ising's model.
 However,  even if the nearest-neighbor correlations of the two models coincide, long-range spatial order
  need not be the same in the two models. 
 In particular, it has been shown that spatial correlations described by $g(r)$   in the Ising's model decrease exponentially fast \cite{Evans1984} in marked contrast with the power-law behavior observed in the Hiebeler's model.
  Moreover, starting from a given Ising's model with parameters $(J,h)$, one can derive the corresponding Hiebeler's parameters $(p,q)$. The correspondence is rather complex, since it may not be unique, especially in the case of symmetry breaking. This case occurs when one crosses the Ising critical point, at $h=0$ and $J_c \simeq 0.44$: then a non-zero magnetization state and its opposite are equally equilibrium states (eq 9).  Conversely, starting from the $(p,q)$ Hiebeler's prescription, one can find corresponding Ising's parameters only for a limited range of $(p,q)$ values.
 
The benefits of comparing the models is twofold. First, it allows to understand better the phases probed by Hiebeler's prescriptions, in the light of what we know from the Ising's phase diagram in the $(J,h)$ space.
Secondly, it allows to probe other kinds of phases in the regions of the $(p,q)$ space that cannot be mapped onto an Ising's model.
 
In the following, we will describe in detail all phases encountered in Hiebeler's model in the cases where a clear correspondence with the Ising's model is possible.

\subsection{Ising model}
\label{subsec:ising}

The transformation $S_i= 2X_i-1$ maps the Boolean variables $X_i$ 
(with values $1$ and $0$, used in the context of the Hiebeler's model) to spin variables (with values $+1$ or $-1$).

Considering an Ising's model with reduced \cite{footnote1} 
 ferromagnetic interaction $J$ and reduced external field $h$, it is described by the Hamiltonian:
\be \label{eq:H}
H([S] \,| \,J,h)=-\sum_ihS_i -\sum_{\langle i, j\rangle} JS_iS_j\ee
where the second sum runs over the pairs $\langle i, j\rangle$ of nearest neighbors. The global distribution (that is, the probability distribution in the space of the configurations $[S]$ on the  lattice):
\be \label{eq:P}
P([S]\, |\, J, h) ={ 1\over Z(J,h) } e^{-H([S] \,| \,J,h)}\ee
where $ Z(J,h)=\sum_{S} e^{-H([S] \,| \,J,h)}$ is the partition function, ensuring a proper normalization of the distribution.
The interaction $J$ can be either ferromagnetic ($J>0$) or antiferromagnetic $(J<0)$.

\subsection{Correspondence between Hiebeler's and Ising's model parameters}

The calculation of $p$ and $q$ as a function of the Ising's parameters can be an involved task. Consider for example the definition of $q$:
$P(X_i=1\, |\,X_j=1 )=P(S_i=1 \, |\, S_j=1)= q$, where $i$ and $j$ are neighbors . One can write:

\be
P(S_i=1 \, |\, S_j=1)=
 \frac{\sum_{S_k} P(S_i=1, S_j=1, S_k)}
{\sum_{S_k} (P(S_i=-1, S_j=1, S_k)+P(S_i=1, S_j=1, S_k))}
\ee

Hence $q$ comes from the following formula:

\be
q=\frac{\sum_{S_k}  e^{J+h+J\sum_{k\neq i,j}S_k}}
{\sum_{S_k} (e^{-J-h-J\sum_{k\neq i,j}S_k}+e^{J+h+J\sum_{k\neq i,j}S_k})}
\ee

Because  of non-vanishing terms in the Boltzmann factors involved in the calculation, we deduce that it is not possible
at this stage  to give an analytical correspondence between $(p,q)$ and $(h,J)$ except in the specific case for which $J = 0$.

In the case where $J=0$, the Hamiltonian is reduced to:
$H([S] \,| \,J=0,h)=-\sum_i hS_i $.
Then, we have for any pair $i,j$ of nearest neighbors:
$P(X_i=1\, |\,X_j=1 )=P(S_i=1 \, |\, S_j=1)= {e^{h}\over e^{h}+e^{-h}}=q$ and 
$P(X_i=0\, |\,X_j=0 )=P(S_i=-1 \, |\, S_j=-1)= {e^{-h}\over e^{-h}+e^{+h}}={1-2p+pq\over 1-p}$
yielding explicit formulae for $q(h)$ and $p(h)$.

\begin{figure}[htbp]
\begin{center}
\includegraphics[width=0.48\columnwidth]{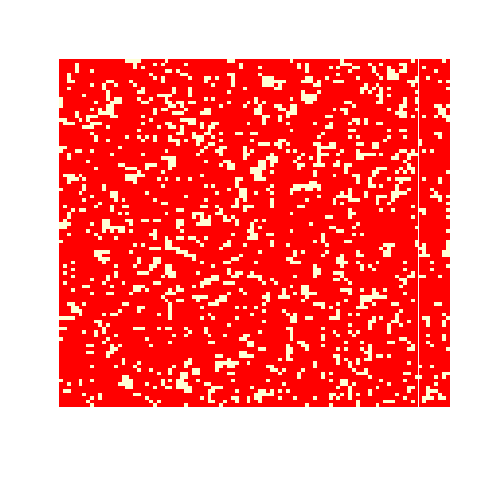}
\includegraphics[width=0.48\columnwidth]{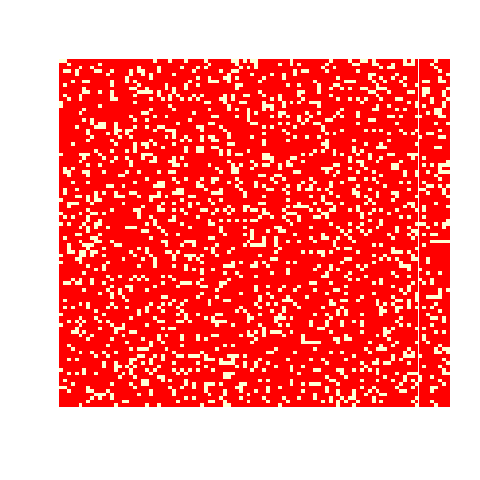}
\caption{(Color online) \ Simulating habitats with Ising's parameters: the two maps have similar $p$ and $q$ and quite different $h$ and $J$. 
Left: $h$ = $J$ = 0.25, $p$ = 0.84, $q$ = 0.87. Right: $h$ = 0.75, $J$ = 0.0, $p$ = $q$ = 0.82. Here, $L=100$.
}
\label{default15}
\end{center}
\end{figure}

This absence of explicit link between $(J,h)$ and $(p,q)$ is due  to the difference of  constraint prescription in the two models.
Starting from $J$ and $h$, it is possible to find the corresponding $p$ and $q$ by numerical simulation, whereas the reverse is not  true: the knwoledge of $p$ and $q$ is not enough 
to determine uniquely $J$ and $h$. As an illustration, one can generate different configurations corresponding to the same values of $(p,q)$ obtained from two very different values of $(J,h)$ (Fig.~\ref{default15}).
   
\subsection{Comparison of Hiebeler's and Ising's phases}

\begin{figure}[htbp]
\begin{center}
\includegraphics[width=0.5\columnwidth]{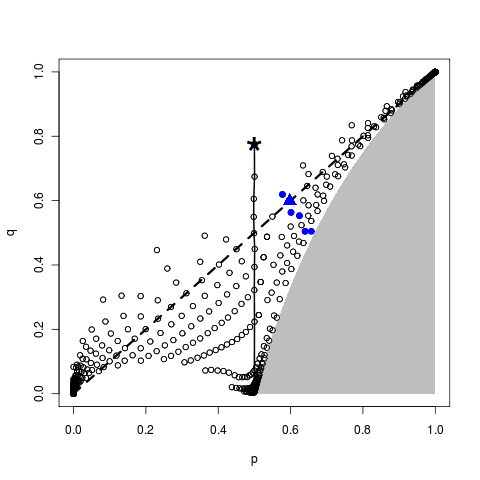}
\caption{(Color online) \ Points of the $(p,q)$ plane reached by an Ising-type simulation. Comparison with Fig.~\ref{default5} shows that a large part of the Hiebeler's phase diagram is not covered.
Remarkable points and lines are visible. Namely: the dashed black diagonal line corresponds to $p=q$ (uncorrelated percolation model), with the uncorrelated percolation threshold 
$p_c=q_c \approx 0.592$ (blue triangle). Blue points are on the correlated percolation line.  The black vertical line corresponds to $p=1/2$, or $h=0$ and $J\leq J_c$; it is a non percolating paramagnetic phase. The line ends at the Ising critical point (black star), which also lies on the correlated percolation transition line; above $J_c$ there is symmetry breaking in $p$ and $q$ (see text).}
\label{default16}
\end{center}
\end{figure}

\begin{figure}[htbp]
\begin{center}
\includegraphics[width=0.5\columnwidth]{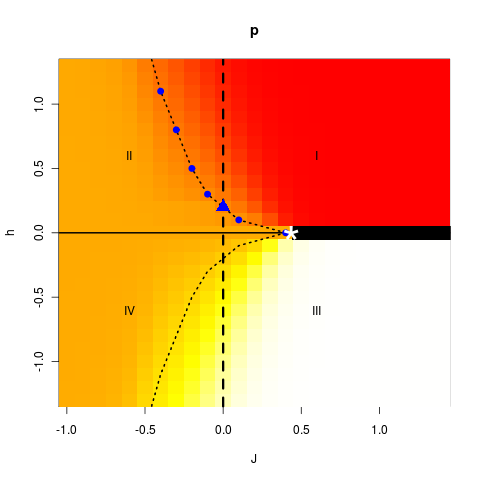}
\includegraphics[width=0.5\columnwidth]{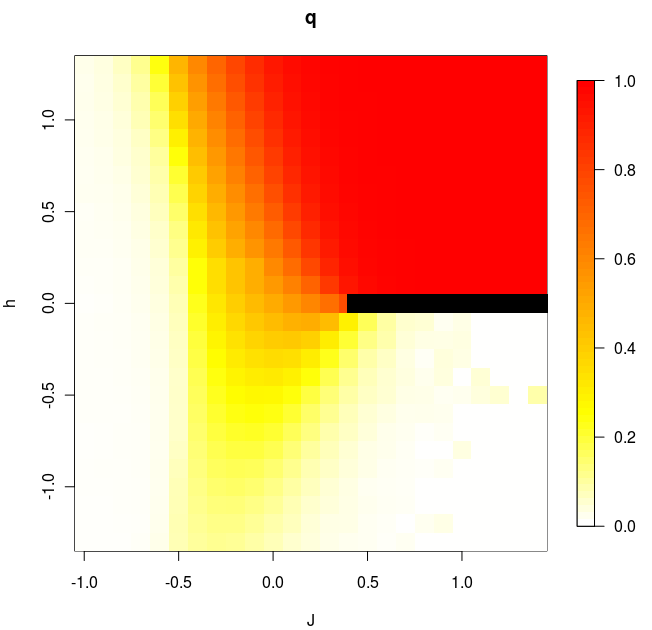}
\caption{(Color online) \ Values of observed Hiebeler's parameters $p$ (top) and $q$ (bottom) in the space of Ising's parameters $(J,h)$. The color code is the same for $p$ and $q$. The four different regions of configurations I, II, III, IV, are described in the text. Remarkable points and lines are visible and can be put in correspondence to those described in Fig.~\ref{default16}.  The black vertical dashed line ($J=0$) corresponds to $p=q$ (uncorrelated percolation model), with the uncorrelated percolation threshold 
$p_c=q_c \approx 0.592$ (blue triangle). The dashed line (and blue points) between regions I and II correspond to the correlated percolation line. The black horizontal line corresponds to $h=0$ and $J\leq J_c$ (or $p=1/2$): it ends at the Ising critical point (white star), which also lies on the correlated percolation transition line. Above $J_c$ there is symmetry breaking in $p$ and $q$ (thick black line).}
\label{default17}
\end{center}
\end{figure}

We have simulated the Ising's model for different values of $(J,h)$ using the Metropolis algorithm.  After the convergence of the algorithm has been reached, we 
calculated the corresponding values of $p$ and $q$   (Figs.~\ref{default16} and \ref{default17}).
Then we tried to relate the phases found for the Ising's model to the phases studied previously in the $(p,q)$ plane.

In Fig.~\ref{default16} and Fig.~\ref{default17} we represent the space of Hiebeler's  parameters $(p,q)$ explored by an Ising numerical simulation. 
We can see that for the (large) amplitude of $J$ and $h$ simulated, only a small part of the space of Hiebeler's model parameters is explored (by comparing Fig.~\ref{default16}
and Fig.~\ref{default5}).  
We also show where the correlation percolation transition line in the $(p,q)$ plane can be found in the $(J,h)$ phase diagram for $p$, but the Ising model does not describe the upper part of this percolation line (for $p<1/2$ and $q$ large).
The in-depth presentation of the correspondence between the two models is provided in detail in the Appendix.

To conclude, we have seen that an Ising model is also an eligible model for correlated percolation, and that it contains part of the percolation phase diagram of the Hiebeler's model.
However, the Hiebeler's model is richer in the sense that it generates other types of configurations. Moreover,  the direct use of the density $p$ and aggregation parameter $q$
are somehow more intuitive in order to construct the desired types of habitat.

We underline that  Ising's model is thoroughly defined by the Hamiltonian $H$, Eq.~\ref{eq:H}, which determines the global probability distribution $P([S])$, Eq.~\ref{eq:P}. In contrast, the knowledge of $p$ and $q$  alone does not determine the global probability distribution.  It is straightforward to show that, in the domain where $(p,q)$ corresponds to a pair $(J,h)$, the least biased distribution  obtained by maximizing the Shannon entropy $-\sum_{[S]} P([S]) \log_2 P([S])$ under the constraints given by  $p$ and $q$ coincides with Ising's distribution \cite{shannon}. This remark  indicates that additional prescriptions are contained in the Hiebeler's algorithm which produce the observed phases and correlation structure. Indeed, the global spatial distribution is not only controlled by the parameters $(p,q)$ but also by the (non unique) way  conditional probabilities at a larger distance and higher moments are related to  $p$ and $q$.
A correspondence between the parameters is not enough to enforce an equivalence between Ising's and Hiebeler's models. 
In conclusion, we found that compared to Ising's model, the Hiebeler's algorithm richness lies in its diversity of types of generated configurations, both inducing longer-range correlations, and designing new configurations not contained in the Ising description.



 \section{Discussion}

 Ecologists often investigate the consequences of highly fragmented habitats in contrasted physical or land-use contexts for the maintenance of biodiversity 
  \cite{Fahrig2003}.
Note that dynamical processes occurring  on the habitat can change drastically the arrangement of suitable sites: in a dynamical model where the suitability of a site modifies back its suitability (mimicking the degradation of a suitable site due to grazing)  \cite{Kefi2007},  a power-law decay  is observed for  a wide range of  $p$ values. Here, we  limited ourselves to the case of a static habitat. We have investigated the properties of the Hiebeler model representing suitable and unsuitable habitat cells in a square lattice. This model is based on a density parameter, $p$, and a parameter of short-range aggregation, $q$. It is a special case of correlated percolation. The habitat parameterization $(p, q)$ is appealing for ecologists dealing with the short-range spatial dynamics of organisms in fragmented habitats, and as such it deserves in depth investigation of its spatial properties. We generated a diversity of habitat maps by varying the parameters $(p, q)$, from fragmented to continuous, from random-like to highly clustered distributions. An important result is that the diversity of habitat configurations generated by the Hiebeler model is richer than the archetypal Ising model, which a statistical physicist would still choose naturally as a correlated percolation model (Figure 16). In fact, the Hiebeler's model encompasses all the Ising phases, namely the antiferromagnetic, ferromagnetic and paramagnetic phases, as well as their transition points. However, we have showed that the space of configurations probed by the model can be different according to the algorithm implemented numerically. We thus underline that the $(p, q)$ parameterization does not provide unambiguous spatial properties, a fact that has been overlooked by ecologists and can strongly influence their models of organism dynamics in fragmented habitats. The parameters $p$ and $q$ are easy to estimate from experimental data. In ecology, habitat density and patch isolation are basic observable quantities that can be estimated from maps of environmental data. However the parameters should represent spatial contexts of habitats in a non ambiguous way, that is, they should fully prescribe the spatial properties. We have shown here that it requires additional information such as algorithmic constraints.   This ambiguity, together with the speed of numerical convergence, are to be taken into account for further ecological studies. Furthermore, we have highlighted other quantities that should be measured by ecologists in order to get more insight on which model to rely on. Remarkably, the pair correlation function $g(r)$ -in contrast to the pair-connectedness $C(r)$- can discriminate between different algorithm implementations of the Hiebeler model. Both quantities are hence of interest in ecology. Note that $C(r)$ is intuitively useful in the framework of an underlying model with short-range dispersal, whereas $g(r)$  is the natural quantity when one is interested by long-range dispersal.


The influence of correlations on the percolation universality class has been previously investigated in the literature, based on the large-distance behavior of $g(r)$. The exponent $\nu$ for the correlation length does not seem to change from its uncorrelated percolation value if the spatial correlations are short-range\cite{Weinrib1984}  \cite{Zhang1982} \cite{ChavesKoiller1995}. The model considered here with $g(r) - p \simeq 1/r$ rather qualifies for long- range correlations. However, our results indicate that its critical properties do not markedly differ from those of the uncorrelated percolation. For all practical purposes, the universality class looks unaffected when non trivial nearest-neighbor conditional probabilities are prescribed (aggregation parameter $q$). Note that in the context of porous materials \cite{Bejan} the Hiebeler model (and is algorithmic variants) may correspond either to equilibrated materials or non-equilibrated materials (in the sense of having different synthesis protocols), all with the same prescriptions $p$ and $q$. These would correspond to structures with the same critical exponents but different spatial  correlations. This would certainly be of interest in future studies of porous materials.

From an ecologist point of view, the Hiebeler's model can be a good one for simulating realistic habitats, as soon as one has evidence that the habitats to be modelled are indeed characterized by short-range aggregation. This can be checked for instance by measuring empirically $P(s)$ and checking that it has an exponent $\tau$  not too far from the value for uncorrelated percolation. Then the aggregation parameter $q$ and the density $p$ can also be measured empirically and Hiebeler's habitats can be simulated with these inputs, with an algorithm consistent with the measured $g(r)$. Even better, if one has a good knowledge of the processes underlying habitat formation, its topology, its ecological constraints, it will be easier to relate them to an instance of Hiebeler's model. This opens the way for further studies of (i) the processes generating these habitats, (ii) the dynamics of organisms living and moving over the fragmented habitats. Although models of population dynamics over suitable localities (metapopulations) were first developed in a mean field perspective on a homogeneous substrate, following the seminal work of Levins \cite{Levins, Lande, Amarasekare}, the colonization and extinction events depend in reality on the spatial distribution of suitable and unsuitable sites. The colonization process depends on the ability of individuals to move from a suitable occupied site to a suitable unoccupied one, is decreasing rapidly with distance, and therefore critically depends on habitat density and aggregation \cite{Ovaskainen2002}. In particular,  persistence will be related to the interplay between the spatial range of colonization and the aggregation parameter of the habitat. Our knowledge of the typical landscapes generated by the Hiebeler  model will hence help us classify the outcome of metapopulation dynamics on such substrates.

\section{Appendix}

In this Appendix, we provide details on the correspondence between Hiebeler's model and Ising's model found in section IV.
We refer in particular to Figures \ref{default16} and \ref{default17} in the main text.



\begin{figure}[htbp]
\begin{center}
\includegraphics[width=0.5\columnwidth]{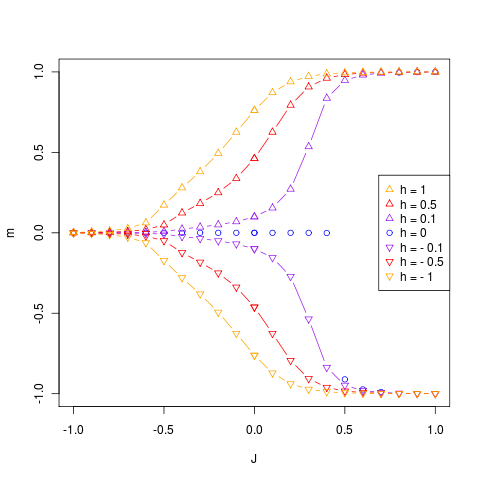}
\caption{(Color online) \ Sketch of the magnetization  profiles as a function of $J$ at fixed $h$, for various values of  $h$.}
\label{default18}
\end{center}
\end{figure}

\begin{figure}[htbp]
\begin{center}
\includegraphics[width=0.5\columnwidth]{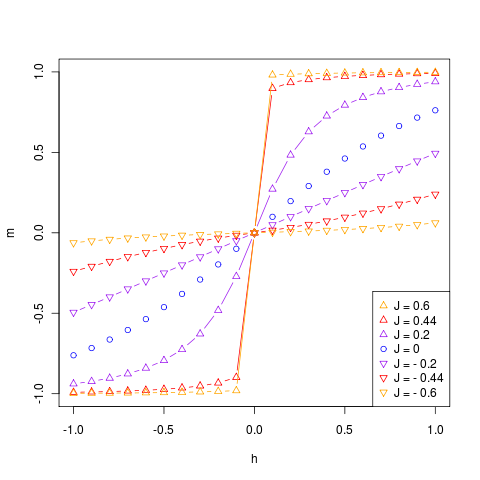}
\caption{(Color online) \ Sketch of the magnetization profiles as a function of $h$ at fixed $J$, for various values of $J$. }
\label{default19}
\end{center}
\end{figure}

The magnetization, defined for spin systems  as $m=1/N \sum_i <S_i>$, where the average is done over different realizations, is related to $p$ by:
$m=2p-1$.  We measured the magnetization as a function of $J$ for various values of $h$ (Fig.~\ref{default18}) and also as a function of $h$ for various values of $J$ 
(Fig.~\ref{default19}).

\begin{figure}[htbp]
\begin{center}
\includegraphics[width=0.5\columnwidth]{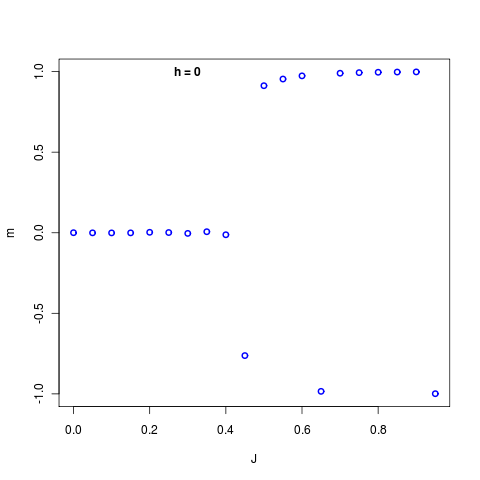}
\includegraphics[width=0.5\columnwidth]{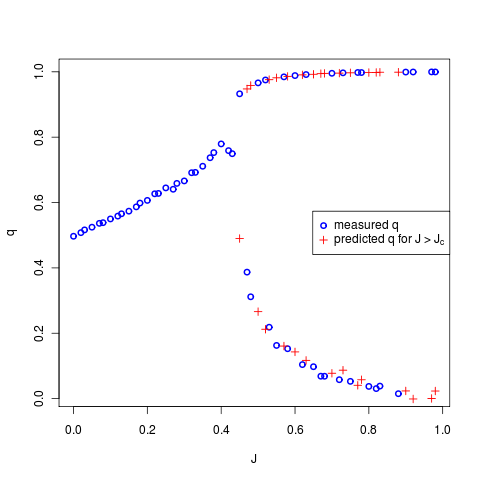}
\caption{(Color online) \ Magnetization $m(J)$ and aggregation parameter $q(J)$ for $h=0$. For $J\geq J_c$ there is a symmetry breaking: the system either jumps to a filled state or to an empty state. 
Hence two values of $m=2p-1$ and $q$ are possible.}
\label{default20}
\end{center}
\end{figure}

\begin{figure}[htbp]
\begin{center}
\includegraphics[width=0.32\columnwidth]{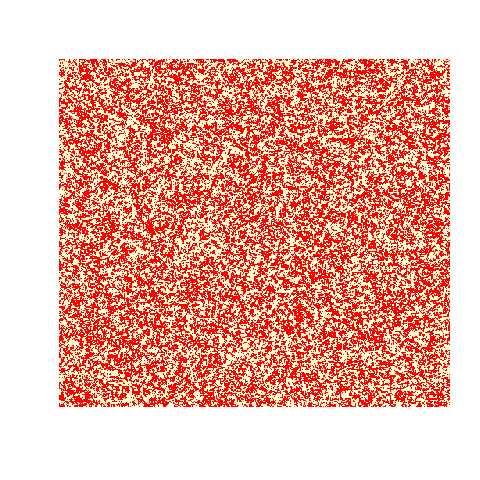}
\includegraphics[width=0.32\columnwidth]{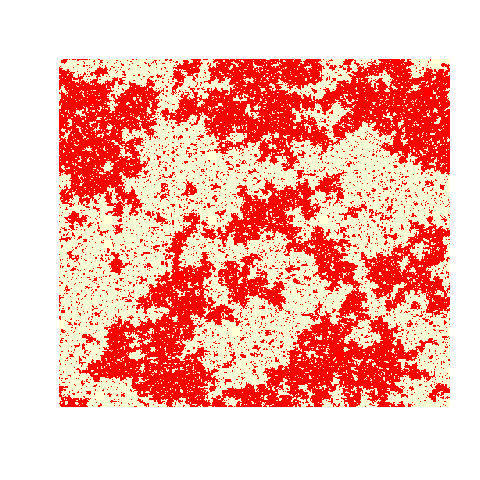}
\includegraphics[width=0.32\columnwidth]{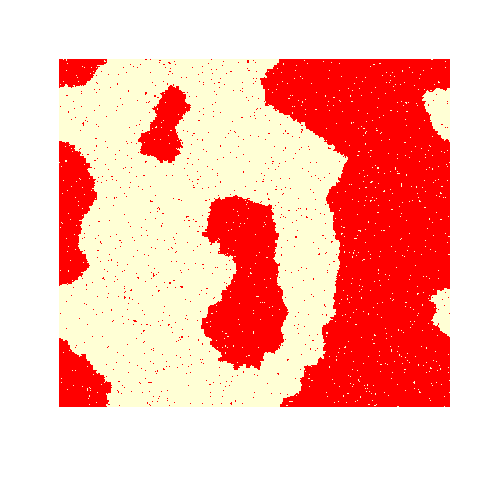}
\caption{(Color online) \ Typical Ising configurations (at $h=0$)
when crossing the Ising critical point. From left to right, $J=0.3, J=J_c=0.44, J=0.6$. Figure 18 can be used to refer to the corresponding $(p,q)$ values. Here $L=500$.}
\label{default21}
\end{center}
\end{figure}

\begin{figure}[htbp]
\begin{center}
\includegraphics[width=0.32\columnwidth]{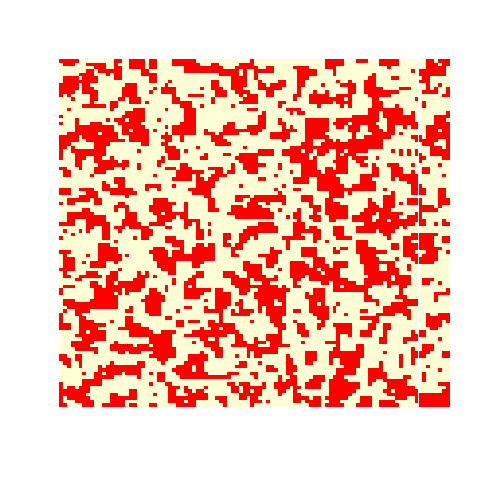}
\includegraphics[width=0.32\columnwidth]{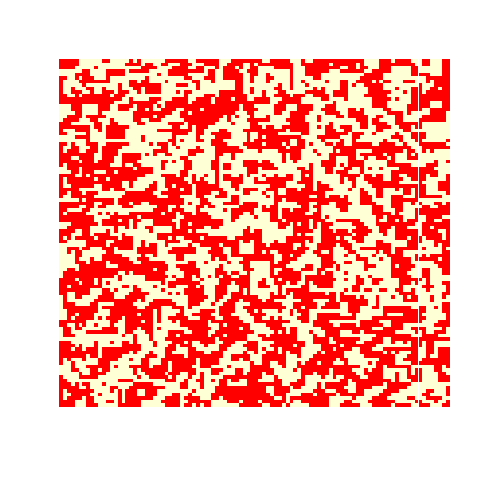}
\includegraphics[width=0.32\columnwidth]{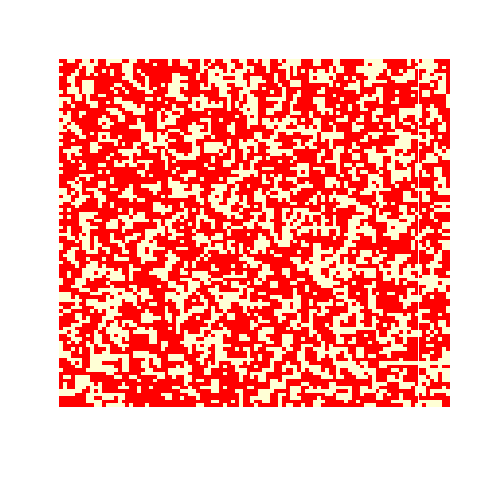}
\caption{(Color online) \ Typical Hiebeler's configurations when crossing the percolation line at $q=0.7$ (see the phase diagram Figure 3): from left to right,
$p=0.4, p=p_c(q=0.7)=0.5223, p=0.7$. Here, $L=100$. Crossing of this critical line manifests in very different spatial landscapes from those observed when crossing the Ising critical line (figure 19).}
\label{default22}
\end{center}
\end{figure}



The usual phase transition of the Ising model can be recovered in this study.
At zero field ($h=0$) and positive $J$, there is a phase transition at $J_c=\ln(1+\sqrt{2})/2 \approx 0.44$. 
This is the well-known critical point of the Ising model, separating the paramagnetic phase ($J<J_c, m=0, p=0.5$) and the ferromagnetic phase 
($J>J_c, m\rightarrow\pm 1, p\rightarrow0,1$).
Above the critical point one can illustrate the symmetry breaking as the density takes the value $p$ or $1-p$ at equilibrium depending on the initial configurations of the simulation. Similarly, two values of the conditional probability can be found: $q$ and  $\frac{1-2p+pq}{1-p}$ (Fig.~\ref{default20}).

\begin{figure}[htbp]
\begin{center}
\includegraphics[width=0.5\columnwidth]{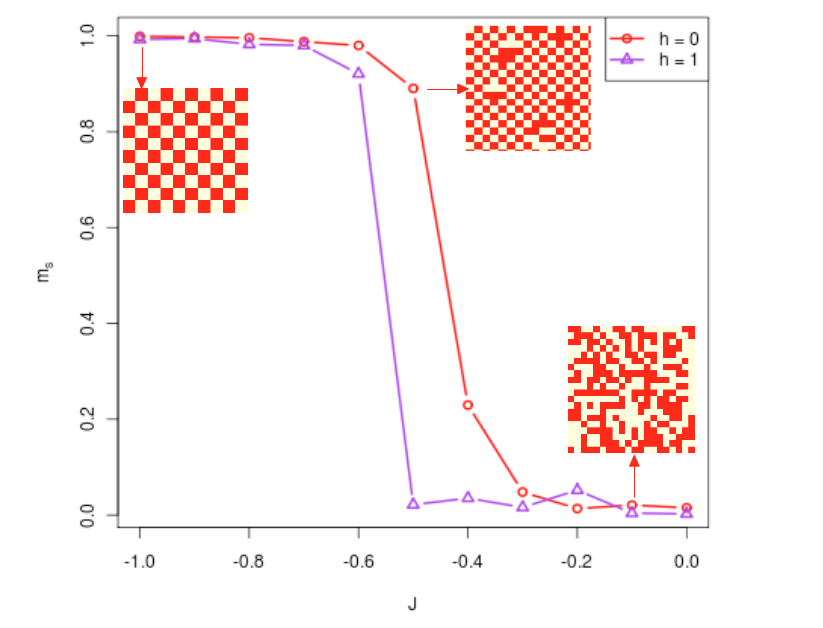}
\caption{(Color online) \ Antiferromagnetic transition: staggered magnetization at fixed  $h=0$ and $h=1$ as a function of $J$. For $h=0$, the transition occurs for $J=-J_c$ (see text).  When a field is applied, antiferromagnetic order sets in
at more negative $J$.}
\label{default23}
\end{center}
\end{figure}

Let us now consider the case where $h=0$ and $ J<0$. It corresponds ---in its larger part--- to values of $p$ close to 1/2 and values of $q$ close to 0.
 It is also well-known that the Ising model undergoes a phase transition at $-J_c\approx -0.44$, where the paramagnetic phase ($J>-J_c$) is separated from the antiferromagnetic phase $(J<-J_c)$. The antiferromagnetic phase is characterized by a zero magnetization, since it corresponds to antiparallel
configurations of spins. However one can characterize this phase by its non-vanishing staggered magnetization $m_s=|m_A-m_B|$. A corresponds to a sub-lattice of lattice unit 2; B is the complementary sub-lattice of lattice unit 2. The antiferromagnetic transition is illustrated in Fig.~\ref{default23} for two different values of $h$.

\begin{figure}[htbp]
\begin{center}
\includegraphics[width=0.5\columnwidth]{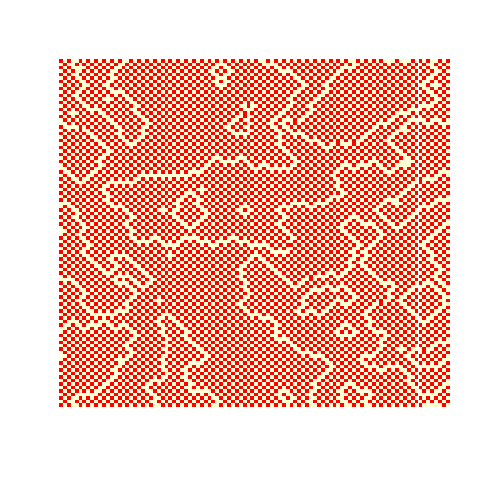}
\caption{(Color online) \ Hiebeler's antiferromagnetic-like configuration, obtained for $p=0.45$ and $q=0.011$ after a very long convergence
time ($10^{11}$ iterations). Here $L=100$.}
\label{default24}
\end{center}
\end{figure}

The antiferromagnetic phase can be found in particular 
in the Hiebeler's phase diagram in a small region close to $q=0$; it is far from the percolating line, and  requires a long time to reach equilibrium (for a typical realization, see
Fig.~\ref{default24}).

We can now  describe systematically the different types of configurations
found by calculating the values of $(p,q)$ in the Ising plane $(J,h)$. 

 In particular, the Ising critical point is on the percolation line of Hiebeler phase diagram. The other points on the percolation line
  correspond to $h\neq 0$: they correspond either to paramagnetic or antiferromagnetic phases.

\begin{itemize}
 
\item
Line $h=0$.

For $J \gg J_c$, one has a ferromagnetic phase, $m=\pm 1$ and $p=0,1$ because there is a symmetry breaking:
according to the initial configuration, simulation generates a mixture of such configurations.
 If $p=0$, $q=0$; if $p=1$, $q=1$. If $J$ is larger but close to $J_c$, in finite size,
 allowed values of $p$ and $q$ can take two possible values:
$(p, 1-p)$ and $(q, \frac{1-2p+pq}{1-p})$.
 
For $-J_c < J < J_c$ there is a  paramagnetic phase, with $m=0$ and $p=1/2$.

For $J=0$, $p=q=1/2$ and then $q$ decreases when $J$ decreases. For $J=J_c$, $q \approx 0.8$
and for $J=-J_c$, $q\approx 0.2$.
 
For $J < -J_c$ there is an antiferromagnetic phase with $p=1/2$. As $J$ decreases, $q$ tends to 0, $m_s$ tends to 1 and a perfectly alternate configuration with suitable and unsuitable sites  is observed (Fig.~\ref{default23}).

\item
Line $J=0$.

In this case there is an exact correspondence
$p=q=\frac{e^h}{e^h+e^{-h}}$. This corresponds in Fig.~\ref{default16} to the diagonal $p=q$, namely uncorrelated percolation model.
Percolation occurs at $p_c=q_c\approx 0.592$, which corresponds to a  value $h_c>0$ such that
$p_c=q_c=\frac{e^{h_c}}{e^{h_c}+e^{-h_c}}$. 
For $h>h_c$ there is percolation, for $h<h_c$ there is no percolation.
 
\item
Transition lines $J_c(h)$ (dashed lines in Fig.~\ref{default17}) correspond either to a paramagnetic/ferromagnetic
 transition if $h=0$ or to a antiferromagnetic/ferromagnetic transition in non-zero
field (cf magnetization and staggered magnetization profiles in Figs.~\ref{default19}  and \ref{default23}).
In fact the transition line $J_c(h)$
 in positive field coincide with the percolation line ($p_c,q_c$) 
of the correlated percolation model (blue dots,  in  both Fig.~\ref{default16} and Fig.~\ref{default17}). 
However, one can have other points on the correlated percolation line (as seen in the Hiebeler's phase diagram, for $p=1/2$ and $q>0.8$) where there is no possible correspondence with an Ising model.
One has also particular points.

\begin{itemize}
\item
the point $p=1/2, q\approx 0.8$ corresponds to the ferromagnetic transition in zero field ($J=J_c, h=0$);
it is a limiting point in the percolation diagram (black star in Fig.~\ref{default16} and white star in Fig.~\ref{default17}).
\item
the point $p=q=0.592..$ corresponds to simple percolation and $J=0,h=h_c$ (blue triangle in Figs.\ref{default16} and \ref{default17}).

\end{itemize}
\end{itemize}

These remarkable points and lines allow to distinguish four different regions I, II, III and IV, that we describe here briefly.
\begin{itemize}
 
\item
Region I: $h>0$, $J>J_c(h)$, the symmetry breaking disappears because of the non-zero field; hence one tends to a ferromagnetic
phase for $J>0$ with $p=1$ and $q=1$. For $J<0$ however, one has an antiferromagnetic phase where $p$ is close to $1/2$ and $q$ varies between 1 and
$1/2$.
 
\item
Region II: $h>0$, $J<J_c(h)$. If $J>0$, one has a paramagnetic phase, with $p=1/2$ and $q$ close to $1/2$.
If $J<0$ one has an antiferromagnetic phase $p=1/2$ and $q=0$.
 
\item
Region III: $h<0$, $J>J_c(h)$, the symmetry breaking disappears because of the non-zero field; hence one tends to a ferromagnetic
phase for $J>0$ with $p=0$ and $q=0$. For  $J<0$ however, one has an antiferromagnetic phase where $p$ is close to $1/2$ 
and $q$ varies between 0 and $0.4$.
 
\item
Region IV: $h<0$, $J<J_c(h)$. If $J>0$, one has a paramagnetic phase, with $p=1/2$ and $q$ close to $1/2$.
If $J<0$ one has an antiferromagnetic phase $p=1/2$ and $q=0$.
 
\end{itemize}


\newpage


\begin{thebibliography}{1000}
\bibitem{footnote1}
``Reduced'' means that we group the temperature factor $\beta=1/k_BT$ with the raw interaction strength into a dimensionless coupling constant $J$.
\end{thebibliography}
\end{document}